\documentclass[a4paper,reqno,12pt]{amsart}
\usepackage[margin=2.5cm,papersize={19.75cm,28cm}]{geometry}
\usepackage[all]{xy}           
\usepackage{amssymb}           
\usepackage{hyperref}
\usepackage{eucal}
\usepackage{epsfig}
\usepackage{graphicx}
\usepackage{color}
\usepackage{graphics}
\usepackage{latexsym}
\usepackage{float}

\numberwithin{equation}{section}

\newtheorem{definition}{Definition}[section]

\newtheorem{theorem}[definition]{Theorem}
\newtheorem{proposition}[definition]{Proposition}

\newtheorem{remarkth}[definition]{Remark}
\newtheorem{example}[definition]{Example}
\newenvironment{remark}{\begin{remarkth}\upshape}{\hfill$\diamond$\end{remarkth}}

\renewcommand{\emph}[1]{{\bfseries\itshape{#1}}}

\newcommand{\R}{\mathbb{R}}      
\newcommand{\F}{\mathbb{F}}
\newcommand{\ra}{\rightarrow}

\newcommand{\lp}{\left(}
\newcommand{\rp}{\right)}

\begin{document}
\title[Higher-order discrete variational problems with constraints]{Higher-order discrete variational problems with constraints}

\author[L.\ Colombo]{Leonardo\ Colombo}
\address{Leonardo Colombo:
Instituto de Ciencias Matem\'aticas (CSIC-UAM-UC3M-UCM), Campus de
Cantoblanco, UAM C/ Nicolas Cabrera, 15 - 28049 Madrid, 28006
Madrid, Spain} \email{leo.colombo@icmat.es}
\author[D.\ Mart\'\i n de Diego]{David Mart\'\i n de Diego}
\address{David Mart\'\i n de Diego:
Instituto de Ciencias Matem\'aticas (CSIC-UAM-UC3M-UCM), Campus de
Cantoblanco, UAM C/ Nicolas Cabrera, 15 - 28049 Madrid, Spain}
\email{david.martin@icmat.es}
\author[M.\ Zuccalli]{Marcela Zuccalli}
\address{Marcela Zuccalli:
Departamento de Matem\'aticas, Universidad Nacional de La Plata,
Calle 50 y 115, 1900 LA Plata, Buenos Aires, Argentina.}
\email{marcezuccalli@gmail.com}

\thanks{This work has been supported by MICINN (Spain) Grant MTM2010-21186-C02-01, MTM 2011-15725-E, ICMAT Severo Ochoa Project SEV-2011-0087 and
IRSES-project "Geomech-246981''. L.C also wants to thank CSIC and
JAE program for a JAE-Pre grant.}



\maketitle


\begin{abstract}
An interesting family of  geometric integrators for Lagrangian systems can be defined using  discretizations of the Hamilton's principle of critical action.
 This family of geometric integrators  is  called variational
integrators.

In this paper, we derive new variational integrators for
higher-order lagrangian mechanical system subjected to higher-order
constraints. From the discretization of the variational principles, we show that our  methods are automatically symplectic and, in consequence, with a very good energy behavior. Additionally, the symmetries of the discrete Lagrangian imply that  momenta is conserved by the integrator. Moreover,  we extend our construction  to variational integrators where the lagrangian is explicitly
time-dependent. Finally,
some motivating applications of higher-order problems are considered; in particular,
optimal control problems for explicitly time-dependent underactuated systems and
an interpolation problem on Riemannian manifolds.

\end{abstract}

\maketitle

\section{Introduction}

\subsection{General background and motivation}

Recently, higher-order variational problems have been studied for
their important applications in aeronautics, robotics,
computer-aided design... where are necessary variational principles
that depend on higher-order derivatives (see
\cite{CeGr,CeIbdLMdD,CoMdD,CoMdDZu,GHMRV10,GHR11,libromanolo}).
 The dynamics of these systems are governed by
 variational principles on higher-order tangent bundles.
Therefore, it is quite interesting to develop  structure-preserving
numerical integration schemes for this kind of systems.

Discrete mechanics has become a field of intensive research activity
in the last decades \cite{welant,mawest,MoVe,veso}. Many of the
geometric properties of a mechanical system in the continuous case
admit an appropriate counterpart in the discrete setting. In this
sense, variational integrators preserve some invariants of the
mechanical system, in particular, momentum and symplecticity (see
\cite{Hair,Kane,Leok,mawest}).

In this paper, we construct a geometric integrator determined  by a
discretization of a variational principle derived by a higher-order
Lagrangian. Such type of discrete mechanical systems have been
recently studied  in \cite{benitoleondediego,BHM, CoJiMdD11}
(without the presence of constraints) for applications in optimal
control, trajectory planning and theoretical physics.

For time stepping algorithms
with fixed time steps, the theorem proved by  Ge and Marsden \cite{Ge}  divides the set of geometric algorithms
 into those that are energy-momentum preserving and those that are
symplectic-momentum preserving. The construction of
energy-momentum-symplectic integrators is indeed possible if one
allows time step
adaptation \cite{Kane}. One purpose of this paper is
to extend the results previously obtained for conservative
mechanical systems with constraints to the case of time-dependent
higher-order lagrangian systems subjected to time-dependent higher-order
constraints following the approach given in \cite{LD1} and  also study
time-dependent higher-order Lagrangian mechanics with either fixed
or adaptive time-stepping.

Some of the possible applications are the following.  The first involves an
important class of controlled mechanical systems,
\textit{underactuated mechanical systems} \cite{bullolewis},
\cite{SeBa} which include spacecraft, underwater vehicles, mobile
robots, helicopters, wheeled vehicles, mobile robots, underactuated
manipulators, etc. The purpose is find a discrete path which solve
the discrete controlled equations obtained by a variational
procedure and minimize a discrete cost function subject to initial
and final boundary conditions.

Another interesting application of  higher-order variational
principle will be \textit{Riemannian cubic splines}  (see
\cite{BHM,leite,GHMRV10,noakes}) which generalizes the typical
Euclidean cubic splines. The problem consists of  minimizing the
mean-square of the covariant acceleration on a Riemannian manifold,
with given initial and final conditions, and also some interpolation
constraints. Many authors call this type of  problems,
\textit{dynamic interpolation problems,} since the trajectories
interpolating the points are obtained through solutions of dynamical
systems, rather than being given a priori by polynomials.  In our
paper, we will propose a  discrete variational method for
interpolating cubic splines on a Riemannian manifold. As an example,
we  consider the discretization of  cubics splines on the sphere
adding holonomic constraint. The restriction from $\R^3$ to the
sphere will give a second-order lagrangian system subjected to a
holonomic constraints,  which is one of the cases studied are in our
paper.

To be self-contained, we first introduce a short  background on variational integration,
discrete mechanics and discrete variational systems with  constraints.

\subsection{Discrete Mechanics and variational integrators}
\label{section2}

Let $Q$ be a  $n$-dimen\-sio\-nal differentiable manifold defining
the configuration space of a lagrangian system. If  we denote by
$(q^i)$ with $1\leq i\leq n$ a local coordinate system on $Q$,
then $(q^i,\dot{q}^i)$ is the  associated local coordinate system
on the tangent bundle $TQ$.

Given a Lagrangian function $L:TQ\rightarrow \R$ that describe the
dynamic of the system, their  trajectories are the solutions of the
Euler-Lagrange
 equations given by
\begin{equation}\label{qwer}
\frac{d}{dt}\left(\frac{\partial L}{\partial\dot
q^i}\right)-\frac{\partial L}{\partial q^i}=0, \quad 1\leq i\leq n.
\end{equation}

It is well known that the origin of these equations is variational
(see \cite{AbMa},\cite{marsden} and references therein) and they are a system of implicit  system of
second order differential equations.

 In the following, we will assume that the
Lagrangian is \textit{regular}; that is, the matrix
$\left(\frac{\partial^2 L}{\partial \dot q^i
\partial \dot q^j}\right)$ is non-singular. Under this regularity
hypothesis, the existence and uniqueness of the solution of the equations is guaranteed.

In order to numerically simulate  these equations, one possibility consists of defining  (see for
example \cite{mawest}) variational integrators  which are derived
from a discrete variational principle. These integrators preserve
the symplectic structure and have a good behavior of the energy of
the system (see \cite{Hair}). In addition, if a symmetry of a Lie
group is considered, they preserve the corresponding momentum.

For discretizing a Lagrangian system, first, it is necessary to replace  the velocity
phase space $TQ$  by the cartesian product $Q \times
Q$ and the lagrangian $L$ by a discrete lagrangian
function $L_d : Q \times Q \rightarrow \mathbb{R}$.

From the discrete Lagrangian $L_d$ we define, for all $N
\in \mathbb{N}$, a discrete action
$\mathcal{A}_{d}:Q^{N+1}\rightarrow\mathbb{R}$ given by
$$  \mathcal{A}_{d}(q_{(0,N)}) :=   \mathcal{A}_{d}(q_{0},q_{1},...,q_{N}):=\sum_{k=1}^{N}L_{d}(q_{k-1},q_{k})$$
where $q_k\in Q$ with $0\leq k\leq N$.

The discrete Hamilton's principle establishes that the solutions of
this system are given by
the extremals of the discrete action given fixed points $q_0$ and
$q_N$.  Extremizing $\mathcal{A}_{d}$ over the space of discrete paths,
$q_{(0,N)}$, with fixed initial and final conditions, we obtain the
\textit{discrete Euler-Lagrange equations}
$$D_{1}L_{d}(q_k, q_{k+1})+D_{2}L_{d}(q_{k-1}, q_k)=0\; , \qquad 1\leq
k\leq N-1,$$ where $D_{1}L_{d}$ and $D_{2}L_{d}$ denotes the derivatives of the discrete lagrangian $L_{d}$ respect to the first and the second argument, respectively.

It is well known that, under some regularity conditions (the
matrix $D_{12}L_d(q_k, q_{k+1})$ is non-singular), it is possible to
define the discrete flow $\Upsilon_d:Q\times Q\rightarrow Q\times
Q$ given by
$$\Upsilon_d(q_{k-1},q_{k}):=(q_{k},q_{k+1})$$
where $q_{k+1}$ is the unique solution of the discrete
Euler-Lagrange equations with initial values $(q_{k-1},q_k)$.

We introduce now  two discrete Legendre transformations
associated to $L_{d}$:
\begin{eqnarray}\nonumber\label{TransLegDiscre}
{\F}^{-}L_{d}:Q\times Q &\rightarrow& T^{*}Q\\\nonumber
\lp q_{0},q_{1}\rp&\mapsto&\lp q_{0},-D_{1}L_{d}\lp q_{0},q_{1}\rp\rp,\\\\\nonumber
{\F}^{+}L_{d}:Q\times Q &\rightarrow& T^{*}Q\\\nonumber
\lp q_{0},q_{1}\rp&\mapsto&\lp q_{1},D_{2}L_{d}\lp q_{0},q_{1}\rp\rp,
\end{eqnarray}
and the discrete Poincar\'e-Cartan 2-form
$\omega_{d}:=\lp{\F}^{+}L_{d}\rp^{*}\omega_{Q}=\lp{\F}^{-}L_{d}\rp^{*}\omega_{Q}$,
where $\omega_{Q}$ is the canonical symplectic form on $T^{*}Q$.
If the discrete Lagrangian $L_{d}$ is regular, that is, the matrix
$\left(\frac{\partial^2 L_d}{\partial q_k\partial q_{k+1}}\right)$
is non-degenerate  then $\omega_d$ is a symplectic form.   These
conditions are also   equivalent  to that ${\F}^{-}L_{d}$ or
${\F}^{+}L_{d}$ are  local diffeomorphisms.

 The discrete algorithm determined by $\Upsilon_{d}$ preserves the
symplectic structure on $(T^{*}(Q\times Q),\omega_d)$, i.e.,
$\Upsilon_{d}^*\omega_d=\omega_d$. Moreover, if $G$ acts on $Q$ and
the discrete Lagrangian is invariant under the diagonal action
associated on $Q\times Q$, then the discrete momentum map $J_d\colon
Q\times Q \to \mathfrak{g}^*$ defined by
\[ \langle
J_d(q_k, q_{k+1}), \xi\rangle:=\langle D_2L_d(q_k, q_{k+1}),
\xi_Q(q_{k+1})\rangle \] is preserved by the discrete flow. Here,
$\xi_Q$ denotes the fundamental vector field determined by $\xi\in
\mathfrak{g}$, where $\mathfrak{g}$ is the Lie algebra of $G$,
$$\xi_{Q}(q)=\frac{d}{dt}\Big|_{t=0} (\exp(t\xi)\cdot q)$$ for
$q\in Q$ (see \cite{mawest} for more details). Therefore, these
integrators are symplectic-momentum preserving.

%
%
%

Now,  consider a lagrangian system with constraints determined
by a constraint submanifold $\mathcal{M}$ of $TQ$ given by the
vanishing of $m$ (independent) differential functions $\phi^{\alpha}
: TQ \rightarrow\mathbb{R}$. If we discretize this system, the
submanifold $\mathcal{M}$ is replaced by a discrete constraint
submanifold $\mathcal{M}_d \subset Q \times Q$ determined by the
vanishing of $m$ independent constraints functions $\phi_d^{\alpha}
: Q \times Q \rightarrow \mathbb{R}$.

In order to find the trajectories of this discrete lagrangian
system with constraints from a variational point of view, we
compute the critical point of a discrete action subjected to
the constraint equations; that is,
\begin{equation}\label{mnb-1}
 \left\{
   \begin{array}{l}
   \hbox{min }\mathcal{A}_{d}(q_{(0,N)})\hspace{1.75 cm}\hbox{with } q_0 \hbox{ and }q_N \hbox{ fixed}\\
   \hbox{suject to }\Phi^{\alpha}_d(q_k,q_{k+1})=0, \hspace{1.cm} 1\leq \alpha\leq m \ \hbox{ and }\ 0\leq k\leq
   N-1\; .
   \end{array}
 \right.
\end{equation}


 We define the \textit{augmented Lagrangian} $\widetilde{L}_d:
Q\times Q\times \R^m\to \R$ by
$$\widetilde{L}_d(q_0,q_1,\lambda) :=
L_d(q_0,q_1)+\lambda_\alpha\Phi^\alpha_d(q_0,q_1).$$ This Lagrangian
gives rise the following unconstrained discrete variational problem,
\begin{equation}\label{mnb-2}
 \left\{
 \begin{array}{l}
  \hbox{min }
\widetilde{\mathcal{A}}_{d}\;(q_{(0,N)},\lambda^{(0,N-1)})\;\hspace{.3 cm }\hbox{ with } q_0 \hbox{ and }
 q_N \hbox{ fixed }\; \\
 q_k\in Q\hspace{.5 cm }\lambda_k\in \R^{m}\hspace{.5 cm }
k=0,\ldots,N-1,\hspace{.5 cm} q_N\in Q \hspace{.5 cm }
 \end{array}
\right.
\end{equation}
where
\begin{eqnarray*}\label{pvdsv}\;\widetilde{\mathcal{A}}_{d}\;(q_{(0,N)},\lambda^{(0,N-1)}):=
\sum_{k=0}^{N-1}\;\widetilde{L}_d(q_k,q_{k+1},\lambda^{k})
\end{eqnarray*}
 and
$\lambda^k$ is a $m$-vector with components $\lambda^k_{\alpha},$ $
1\leq \alpha\leq m$, which plays the roll of the lagrangian multipliers.

From the classical lagrangian multiplier lemma and under some regularity conditions, its well know that
the solutions of Problem \eqref{mnb-1} are the same that the ones in
Problem \eqref{mnb-2}.
  Therefore, applying standard discrete variational calculus we deduce that the solutions of problem \eqref{mnb-1} verify the following set of
  difference equations
  \begin{equation}\label{lelo1}
\left\{
\begin{array}{l}
D_1L_d(q_k, q_{k+1})+D_2L_d(q_{k-1},q_k) +\\
\lambda_{\alpha}^{k} D_1\Phi^{\alpha}_d(q_k, q_{k+1}) +
\lambda_{\alpha}^{k-1} D_2\Phi^{\alpha}_d (q_{k-1}, q_k) = 0\ \ \
\ \
{1\leq k\leq N-1} \, ,\\
\Phi^{\alpha}_d(q_k,q_{k+1}) = 0\; \  \hspace{0.4 cm}1\leq
\alpha\leq m \; \hspace{0.1 cm} \ \hbox{ and }  \ \ 0\leq k\leq
N-1.
\end{array}
\right.\end{equation}

If the matrix \[\left(
                  \begin{array}{cc}
                    D_{12} L_d+\lambda_{\alpha}D_{12}\Phi^{\alpha}_d & D_2\Phi^{\alpha}_d \\
                    \left(D_1\Phi^{\alpha}_d\right)^T & \mathbf{ 0 }_{m\times m} \\
                  \end{array}
                \right)\] is non-singular, by a direct application of the implicit function
theorem, we deduce that there exists an application \[
\begin{array}{rrcl}
\widetilde{\Upsilon}_d:& {\mathcal M}_d\times \R^m&\longrightarrow& {\mathcal
M}_d\times \R^m,
\end{array}
\]
given by
$\widetilde\Upsilon_d(q_{k-1},q_k,\lambda^{k-1}):=(q_k,q_{k+1},\lambda^{k})
$ where $(q_{k+1},\lambda^{k})$ is the unique solution of equation
$\eqref{lelo1}$ given  $(q_{k-1}, q_{k}, \lambda^{k-1})$.

In \cite{benitoleondediego}, it is shown that the discrete flow
$\widetilde{\Upsilon}_d$ preserves a symplectic form naturally
defined on ${\mathcal M}_d\times \R^m$. Moreover, if $L_d$ and the
constraint $\Phi_d^{\alpha}$ are invariant under the action of a
symmetry Lie group, $\widetilde{\Upsilon}_d$ preserves the associated momentum.

\subsection{Organization of the paper}

The paper is structured as follows. In Section \ref{section3} we
present some variational problems with constraints which will be
later analyzed using the techniques  developed  in Section
\ref{section4}. The first one is an optimal control problem for
underactuated mechanical systems and the second one is an
interpolation problem on a Riemannian manifold.

In Section \ref{section4} we develop a  discrete variational
calculus for  higher-order lagrangian mechanical systems with
higher-order constraints and next, in Section \ref{section5} we
apply these techniques to higher-order discrete time-dependent
Lagrangian systems. Moreover, we construct the theory of discrete
time-dependent second-order constrained systems  with fixed
time-stepping.

Finally, we solve an optimal control problem for an underactuated
time-dependent mechanical systems and an interpolation problem on
Riemannian manifolds using the integrator proposed in Section
\ref{section4}. In this application, cubic splines are restricted
to the sphere introducing holonomic constraints.

\section{Some higher-order variational problems with constraints}
\label{section3}

In this section  we will introduce some notions about
higher-order tangent bundle geometry.

Given the manifold  $Q$, it is
possible to introduce an equivalence relation
 in the set $C^{k}(\R, Q)$ of $k$-differentiable
curves from $\R$ to $Q$. By definition,  two  curves $\gamma_1(t)$ and $\gamma_2(t)$ in $Q$
where $t\in (-a, a)$ with $a\in \R,$
have contact of order  $k$ at $q_0 = \gamma_1(0) = \gamma_2(0)$ if
there is a local chart $(\varphi, U)$ of $Q$ such that $q_0 \in U$
and
$$\frac{d^s}{dt^s}\left(\varphi \circ \gamma_1(t)\right){\big{|}}_{t=0} =
\frac{d^s}{dt^s} \left(\varphi
\circ\gamma_2(t)\right){\Big{|}}_{t=0}\; ,$$ for all $s = 0,...,k.$

The equivalence class of a  curve $\gamma$ will be denoted by
$[\gamma ]_0^{(k)}.$ The set of equivalence classes will be
denoted by $T^{(k)}Q$
and one can see that it has a natural structure of differentiable
manifold. Moreover, $ \tau_Q^k  : T^{(k)} Q \rightarrow Q$ given
by $\tau_Q^k \left([\gamma]_0^{(k)}\right) = \gamma(0)$ is a fiber
bundle called the \textit{tangent bundle of order $k$} of $Q.$

Given a differentiable function $f: Q\longrightarrow \R$ and $l \in
\{0,...,k\}$, its $l$-lift $f^{(l, k)}$ to $T^{(k)}Q$, $0\leq l\leq
k$, is the
differentiable
function defined as
\[
f^{(l, k)}([\gamma]^{(k)}_0)=\frac{d^l}{dt^l}
\left(f \circ \gamma(t)\right){\Big{|}}_{t=0}\; .
\]
Of course, these definitions can be applied to functions defined on
open sets of $Q$.

From a local chart $(q^i)$ on a neighborhood $U$ of $Q$, it is
possible to induce local coordinates
 $(q^{(0)i},q^{(1)i},\dots,q^{(k)i})$ on
$T^{(k)}U=(\tau_Q^k)^{-1}(U)$, where $q^{(s)i}=(q^i)^{(s,k)}$ if
$0\leq s\leq k$. Sometimes, we will use the standard  conventions,
$q^{(0)i}\equiv q^i$, $q^{(1)i}\equiv \dot{q}^i$ and $q^{(2)i}\equiv
\ddot{q}^i$.

In this section we present two interesting higher-order
variational problems that we  will be study in this paper: an
underactuated optimal control problem and an interpolation problem
on Riemannian manifolds.

\subsection{Optimal control for underactuated mechanical systems}\label{asw}

Consider an underactuated Lagrangian control systems; that is, a
Lagrangian control system such that the number of the control inputs
is less than the dimension of the configuration space
(superarticulated mechanical system following the nomenclature
introduced in \cite{Ba}) $Q$ which is the cartesian product of two
differentiable manifolds $Q = Q_1\times Q_2$. Denote by $(q^A)=(q^a,
q^{\alpha})$ with $1\leq A\leq n$ a coordinate local system on $Q$,
where $(q^a)$ $ (1\leq a\leq r)$ and $(q^{\alpha})$ $(r+1\leq
\alpha\leq n)$ are local coordinates on $Q_1$ and $Q_2$
respectively. In what follows we assume that all control systems are
controllable; that is, for any two points $x_0$ and $x_f$ in the
configuration space, there exits and admissible control $u(t)$
defined on some interval $[0,T]$ such that the system with initial
condition $x_0$ reaches the point $x_f$  at time $T$ (see for more
details \cite{Blo,bullolewis}).

 Adding the control subset $U\subset\R^r$ where $u(t)\in U$ is
the control parameter. We assume that the controlled external
forces $(u^a)$ can be applied only on $Q_1$.

Thus, given the Lagrangian $L:TQ=TQ_1\times
TQ_2\rightarrow\mathbb{R},$ the motion equations of the system are
written as
 \begin{equation}\label{lagrange con control}
\begin{split}
&\frac{d}{dt}\left(\frac{\partial L}{\partial \dot q^a}\right)-
\frac{\partial L}
{\partial q^a}=u^a \\
&\frac{d}{dt}\left(\frac{\partial L}{\partial \dot q^\alpha}\right)-
\frac{\partial L}{\partial q^\alpha}=0\,
\end{split}
\end{equation}
where $a = 1,\ldots,r$ and $ \alpha = r+1,\ldots,n.$

Given a cost function $C: TQ_1 \times TQ_2 \times U \rightarrow
\R$, the optimal control problem consists on finding a trajectory
$(q^a(t), q^{\alpha}(t), u^a(t))$ of state variables and control
inputs satisfying equations \eqref{lagrange con control} from
given initial and final conditions

$(q^{a}(t_0), q^{\alpha}(t_0), \dot{q}^{a}(t_0),
\dot{q}^{\alpha}(t_0))$ and $(q^{a}(t_f), q^{\alpha}(t_f),
\dot{q}^{a}(t_f), \dot{q}^{\alpha}(t_f))$ respectively, minimizing
the cost functional
\[
{\mathcal A}(q(\cdot)):=\int_{t_0}^{t_f} C(q^a, q^{\alpha},
\dot{q}^a, \dot{q}^{\alpha}, u^a)\, dt.
\]

It is well know (see \cite{Blo}) that this optimal control problem
is equivalent to the following second-order variational problem
with second-order constraints:

Extremize
\[ \overline{\mathcal A}(q(\cdot)):=\int_{t_0}^{t_f}
\overline{L}(q^a(t), q^{\alpha}(t), \dot{q}^a(t),
\dot{q}^{\alpha}(t), \ddot{q}^a(t), \ddot{q}^{\alpha}(t))\, dt
\]
subject to the second order constraints given by
\[
\Phi^{\alpha}(q^a, q^{\alpha}, \dot{q}^a, \dot{q}^{\alpha},
\ddot{q}^a, \ddot{q}^{\alpha}):=\frac{d}{dt}\left(\frac{\partial
L}{\partial \dot q^{\alpha}}\right)- \frac{\partial L}{\partial
q^{\alpha}}=0\;  \hbox{ with } \alpha=r+1,...,n
\]
where $\overline{L}:T^{(2)}Q\rightarrow\mathbb{R}$ is defined as
\[
\overline{L}(q^a, q^{\alpha}, \dot{q}^a, \dot{q}^{\alpha},
\ddot{q}^a, \ddot{q}^{\alpha}):= C\left(q^a, q^{\alpha},
\dot{q}^a, \dot{q}^{\alpha}, \frac{d}{dt}\left(\frac{\partial
L}{\partial \dot q^a}\right)- \frac{\partial L}{\partial
q^a}\right).
 \]

Thus, a second order variational problem can be used for reformulate
this type of  underactuated optimal control problem. For more
details about this problem see \cite{CoMdD} and \cite{CoMdDZu} for
the case when the configuration space is a Lie group.

\subsection{Interpolation problem on Riemannian
manifolds}\label{interpolation}

The construction of interpolating splines on manifolds is useful in
many applications (see \cite{leite, GHMRV10,husbloch1,noakes}).
Consider a Riemannian manifold $(Q,\mathcal{G})$ where $\mathcal{G}$
is the metric and $\frac{D}{Dt}$ is the covariant derivative
associated to the Levi-Civita connection $\nabla$. If $(q^i)$ is a
local coordinate system on $Q$, the covariant derivative of the
velocity $\dot{q}$ is locally given by
$$\frac{D}{Dt}\dot{q}=\ddot{q}^{k}+\Gamma_{ij}^{k}(q)\dot{q}^{i}\dot{q}^{j}$$ where $\Gamma_{ij}^{k}(q)$
are the Christoffel symbols of the metric $\mathcal{G}$ at point
$q$.

Then, one can consider the Lagrangian $L:T^{(2)}Q\ra\R$ defined as
\begin{equation}\label{lagrangian1}
L(q,\dot{q},\ddot{q}):=\frac{1}{2}\mathcal{G}_
q\left(\frac{D}{Dt}\dot{q},\frac{D}{Dt}\dot{q}\right)
\end{equation}

 Given
$N+1$ points $q_i\in Q$ with $i=0,\ldots,N$ and tangent vectors
$v_0 \in T_{q_0}Q$ and  $v_N \in T_{q_N}Q,$ the interpolation
problem consists of finding a curve which minimize the action,
\begin{equation}\label{action1}
\mathcal{A}(q(\cdot))=\int_{t_0}^{t_N} L(q,\dot{q},\ddot{q})
dt
=\frac{1}{2}\int_{t_0}^{t_N}\mathcal{G}_{q(t)}\left(\frac{D}{Dt}\dot{q}(t),
\frac{D}{Dt}\dot{q}(t)\right)dt,
\end{equation} among all the  continuous curves defined on $[t_0,t_N],$ smooth on $[t_i,t_{i+1}],$
for $t_0\leq t_1\leq\ldots\leq t_N$, subject to the interpolating
constraints $$q(t_i)=q_i \ \  \mbox{for all} \ \ i\in
\{2,\ldots,N-2\}$$ and the boundary conditions
$$q(t_0)=q_0,\quad q(t_N)=q_N,$$
$$\frac{Dq}{dt}(t_{0})=v_{0},\quad \frac{Dq}{dt}(t_N)=v_N.$$

It is possible to extend  this problem to higher-order systems,
called \textit{higher-order Riemannan splines}. In this case, we
may consider the lagrangian $L_k:T^{(k)}Q\ra\R$ given by

$$L_k(q,\dot{q},\ldots,q^{(k)}):=\frac{1}{2}\mathcal{G}\left(\frac{D^{k-1}}{Dt^{k-1}}\dot{q},\frac{D^{k-1}}
{Dt^{k-1}}\dot{q}\right),$$ for $k>2$ (see \cite{GHMRV10}) where
$\displaystyle{\frac{D^{k-1}}{Dt^{k-1}}}$ denotes the $k-1$
covariant derivative associated to the connection $\nabla$.

Given $N+1$ points $q_i\in Q$ with $i=0,\ldots,N$ and tangent
vectors $v_0^{(l)}\in T_{q_0}^{(l)}Q$ and $v_N^{(l)}\in
T_{q_{N}}^{(l)}Q$, the higher-order interpolation problem consists
of minimizing the action

$$\mathcal{A}(q(\cdot)):= \int_{t_0}^{t_N} L_k(q,\dot{q},\ldots,q^{(k)}) dt  =\frac{1}{2}\int_{t_0}^{t_N}
\mathcal{G}\left(\frac{D^{k-1}}{Dt^{k-1}}\dot{q},\frac{D^{k-1}}{Dt^{k-1}}\dot{q}\right)dt,$$
where the curves $q(t)\in Q,$ are  continuous in $[t_0,t_N]$ and $k-1$
piecewise smooths on $[t_i,t_{i+1}],$ for $t_0\leq
t_1\leq\ldots\leq t_N$ subjected to the interpolation constraints
$$q(t_i)=q_i \ \ \mbox{for all} \ \ i\in \{2,\ldots,N-2\}$$
and the $2k$
boundary conditions
$$q(t_0)=q_0,\quad q(t_N)=q_N,$$
$$\frac{D^{(l)}q}{dt^l}(t_{0})=v^{(l)}_{0},\quad \frac{D^{(l)}q}{dt^{l}}(t_N)=v^{(l)}_N.$$
for all $1\leq l\leq k-1.$

Thus, the Euler-Lagrange equations for the higher-order
Lagrangians $L_k$ are given by

$$\frac{D^{2k-1}}{Dt^{2k-1}}\dot{q}(t)+\sum_{j=2}^{k}(-1)^{j}R\left(\frac{D^{2k-j-1}}{Dt^{2k-j-1}}\dot{q}(t),\frac{D^{j-2}}{Dt^{j-2}}\dot{q}(t)\right)\dot{q}(t)=0,$$
where $R$ denotes the curvature tensor associated to $\nabla$ (see
\cite{CaLeCr,CaLeCr2,noakes}).

\section{Higher-order algorithm for variational calculus with
higher-order constraints}\label{section4}

In this section an integrator for higher-order mechanics with
higher-order constraints is derived from a discrete variational
principle by considering some regularity condition. We show that
this algorithm preserves the discrete symplectic structure and the
momentum associated to a Lie group of symmetries.

\subsection{Higher-order discrete variational calculus}
The natural space substituting the higher-order tangent bundle
$T^{(k)}Q$ is $Q^{k+1}$ (the cartesian product of $k+1$-copies of $Q$) and therefore a higher-order discrete
Lagrangian is an application $L_d:Q^{k+1}\rightarrow\mathbb{R}.$
For simplicity, we use the notation as in
\cite{benitoleondediego}: if $(i,j)\in (\mathbb{N}^{*})^2$ with $
i<j$, $q_{(i,j)}$ denotes the $(j-i)+1$-upla
$(q_i,q_{i+1},...,q_{j-1},q_j).$

Fixed initial and final conditions $(q_{(0,k-1)},q_{(N-k+1,N)})\in
Q^{2k}$ with $N>2k,$ we define the set of admissible curves with
boundary conditions $q_{(0,k-1)}$ and $q_{(N-k+1,N)}$
$$C^{N}(q_{(0,k-1)},q_{(N-k+1,N)}):=\{\overline{q}_{(0,N)}\mid\overline{q}_{(0,k-1)}=q_{(0,k-1)},\overline{q}_{(N-k+1,N)}=q_{(N-k+1,N)}\}.$$

Let us define the discrete action over an admissible sequence
discrete path as
$\mathcal{A}_{d}:C^{N}(q_{(0,k-1)},q_{(N-k+1,N)})\rightarrow\mathbb{R}$
given by
$$ \mathcal{A}_{d}(q_{(0,N)}):= {\displaystyle\sum_{i=0}^{N-k}L_d(q_{(i,i+k)})}.$$

The discrete variational principle states that the solutions of the
discrete system determined by $L_d$ must extremize the action on
the curves with given fixed points. Thus, we obtain the following
system of $(N-2k+1)n$ difference equations.
\begin{eqnarray}\label{eq11}
D_{k+1}L_d(q_{(0,k)})+...+D_1L_d(q_{(k,2k)})&=&0,\nonumber\\
D_{k+1}L_d(q_{(1,k+1)})+...+D_1L_d(q_{(k+1,2k+1)}&=&0,\nonumber\\
...&=&0,\\
D_{k+1}L_d(q_{(N-2k,N-k)})+...+D_1L_d(q_{(N-k, N)})&=&0.\nonumber
\end{eqnarray}

Here, given a smooth function $F:Q^{k+1}\ra\R,$ $D_{j}F$ denotes
the derivative on the $j$-factor of $F.$

These equations are called \textit{higher-order discrete
Euler-Lagrange equations.} Under some regularity hypotheses it is
possible to define a discrete flow $\Upsilon_d:Q^{2k}\rightarrow
Q^{2k}$ by $$\Upsilon_{d}(q_{(i,2k+i-1)}):=q_{(i+1,2k+i)}$$ from
equations \eqref{eq11}. In \cite{benitoleondediego} the authors
proof that this flow is symplectic-momentum preserving.

\subsection{Higher-order algorithm for variational calculus with
higher-order constraints} In this subsection we consider a
higher-order Lagrangian systems with higher-order constraints
given by $m$ smooth (independent) functions
$\Phi^{\alpha}_{d}:Q^{k+1}\rightarrow\mathbb{R} $ with
$1\leq\alpha\leq m.$

 We denote by
$\widetilde{\mathcal{M}}_{d}$ the constraints submanifold of
$Q^{2k}$ locally determined by the vanishing of these $m$
functions. Then,

$$\widetilde{\mathcal{M}}_{d}:=\{q_{(i,i+k)}\mid \Phi^{\alpha}_{d}(q_{(i,i+k)})=0 \hbox{ where }
1\leq\alpha\leq m \hbox{ and } 0\leq i\leq N-k\}.$$

Therefore, we can consider the following problem called
\textit{higher-order discrete variational calculus with
constraints}
$$\left\{
                                       \begin{array}{ll}
                                         \min \mathcal{A}_{d}(q_{(0,N)}) \hbox{ with } (q_{(0,k-1)}, q_{(N-k+1,N)}) \hbox{ fixed } & \\
                                         \hbox{ subject to } \Phi_{d}^{\alpha}(q_{(i,i+k)})=0 \hbox{ with } 1\leq\alpha\leq m \hbox{ and } 0\leq i\leq N-k. &
                                       \end{array}
                                     \right.$$

 It is well know that this classical optimization problem with
higher-order constraints is equivalent to the following
unconstrained higher-order variational problem (which results
singular) for
$\widetilde{L}_{d}(q_{(i,i+k)},\lambda_{\alpha}^{i}):=L_{d}(q_{(i,i+k)})+\lambda^{i}_{\alpha}\Phi_{d}^{\alpha}(q_{(i,i+k)})$
defined on $Q^{k+1}\times\mathbb{R}^{m}$ with $q_{(i,i+k)}\in
Q^{k+1},$
$(\lambda_{\alpha})=(\lambda_{1},...,\lambda_{m})\in\mathbb{R}^{m},
\hbox{ } 0\leq i\leq N-k:$

$$\left\{
                                       \begin{array}{ll}
                                         \min \widetilde{\mathcal{A}}_{d}(q_{(0,N)}, \lambda^{(0,N-k)}) \hbox{ with } (q_{(0,k-1)}, q_{(N-k+1,N)}) \hbox{ fixed } & \\
                                          q_{(i,i+k)}\in Q^{k+1}\hbox{ and } \lambda^{i}\in\mathbb{R}^{m} \hbox{ with } 0\leq i\leq N-k &
                                       \end{array}
                                     \right.$$
where
\begin{equation}\label{action3}\widetilde{\mathcal{A}}_{d}(q_{(0,N)},\lambda^{(0,N-k)}):=\sum_{i=0}^{N-k}\widetilde{L}_{d}(q_{(i,i+k)},\lambda_{\alpha}^{i}),
\end{equation}
$\lambda^{(0, N-k)} := (\lambda^0,...,\lambda^{N-k})$ and
$\lambda^{i}$ is a vector with components $\lambda_{\alpha}^{i},
1\leq\alpha\leq m$.

In the next, we do not  impose the boundary conditions
$(q_{(0,k-1)}, q_{(N-k+1,N)}).$ Thus, we consider as space of
admissible paths
$$C^{(N,N-k)}:=\{(q_0,q_1,...,q_N,\lambda^0,\lambda^1,...,\lambda^{N-k})\in
Q^{N+1}\times\mathbb{R}^{(N-k)m}\},$$ and computing the
differential of the action
\begin{eqnarray}
&&d\widetilde{\mathcal{A}}_{d}(q_{(0,N)},\lambda^{(0,N-k)})\cdot(\delta q_{(0,N)},\delta\lambda^{(0,N-k)})=\nonumber\\
&&\sum_{i=0}^{k-1}\left(\sum_{j=1}^{i+1}D_{j}L_{d}(q_{(i-j+1,i-j+1+k)})+\lambda_{\alpha}^{i-j+1}D_{j}\Phi_{d}^{\alpha}(q_{(i-j+1,i-j+1+k)})\right)\delta q_{i}+\nonumber\\
&&\sum_{i=k}^{N-k}\left(\sum_{j=1}^{k+1}D_{j}L_{d}(q_{(i-j+1,i-j+1+k)})+\lambda_{\alpha}^{i-j+1}D_{j}\Phi_{d}^{\alpha}(q_{(i-j+1,i-j+1+k)})\right)\delta q_{i}+\nonumber\\
&&\sum_{i=N-k+1}^{N}\left(\sum_{j=i-N+k+1}^{k+1}D_{j}L_{d}(q_{(i-j+1,i-j+1+k)})\right.\\
&&\left.+\lambda_{\alpha}^{i-j+1}D_{j}\Phi_{d}^{\alpha}(q_{(i-j+1,i-j+1+k)})\right)\delta q_{i}+ \sum_{i=0}^{N-k}\Phi_{d}^{\alpha}(q_{(i,i+k)})\delta\lambda_{\alpha}^{i}\nonumber.
\end{eqnarray}

The two expressions corresponding to the boundary terms are called
the \textit{Discrete Poincar\'e-Cartan} 1-forms on
$Q^{2k}\times\mathbb{R}^{km}$ and they are given by

\begin{eqnarray*}
&&\Theta_{\widetilde{L}_{d}}^{-}(q_{(0,2k-1)},\lambda^{(0,k-1)}):=\nonumber\\
&&-\sum_{i=0}^{k-1}\left(\sum_{j=1}^{i+1}D_{j}L_{d}(q_{(i-j+1,i-j+1+k)})+\lambda_{\alpha}^{i-j+1}D_{j}\Phi_{d}^{\alpha}(q_{(i-j+1,i-j+1+k)})\right) dq_{i}\nonumber\\
\end{eqnarray*} and
 \begin{eqnarray*}
&&\Theta_{\widetilde{L}_{d}}^{+}(q_{(0,2k-1)},\lambda^{(0,k-1)}):=\nonumber\\
&&\sum_{i=N-k+1}^{N}\left(\sum_{j=i-N+k+1}^{k+1}D_{j}L_{d}(q_{(i-j+1,i-j+1+k)})+\lambda_{\alpha}^{i-j+1}D_{j}\Phi_{d}^{\alpha}(q_{(i-j+1,i-j+1+k)})\right) dq_{i}.\nonumber\\
\end{eqnarray*}

In order to write the higher-order discrete Euler-Lagrange equations
in an analogous way to discrete Euler-Lagrange equations according
to \cite{mawest} we may define the discrete higher-order
Euler-Lagrange operator
$\mathcal{E}\widetilde{L}_d:Q^{2k+1}\times\mathbb{R}^{(N-k)m}\rightarrow
T^{*}Q^k$ given by
\begin{eqnarray*}&&\mathcal{E}\widetilde{L}_{d}(q_{(i,2k+i)},\lambda^{(i,N-k+i-1)}):=\\
&&\sum_{j=1}^{k+1}\left[D_jL_{d}(q_{(i-j+1+k,i-j+1+2k)})+\lambda_{\alpha}^{i-j+k+1}D_j\Phi_{d}^{\alpha}(q_{(i-j+k+1,i-j+2k+1)})\right]dq_{i+k}.
\end{eqnarray*}

%
Summarizing, we have the following result

\begin{theorem}

If $L_d:Q^{k+1}\rightarrow\mathbb{R}$ is a discrete Lagrangian and
$\Phi_{d}^{\alpha}:Q^{k+1}\rightarrow\mathbb{R}$ with $1\leq\alpha\
m$ are $m$ (independent) smooth functions, there exists a unique
differential mapping
$\mathcal{E}\widetilde{L}_{d}:Q^{2k+1}\times\mathbb{R}^{(N-k)m}\rightarrow
T^{*}Q^k$ and there exist two 1-forms
$\Theta_{\widetilde{L}_{d}}^{+}$ and
$\Theta_{\widetilde{L}_{d}}^{-}$ on $Q^{2k}\times\mathbb{R}^{km},$
such that for all variations $(\delta q_{0},...,\delta q_{N})$ and
$(\delta\lambda_{\alpha}^{0},...,\delta\lambda_{\alpha}^{N-k})$ the
differential of the discrete action $\widetilde{\mathcal{A}}_{d}$
defined in \eqref{action3} verifies the following equality
\begin{eqnarray*}
d\widetilde{\mathcal{A}}_{d}(q_{(0,N)},\lambda^{(0,N-k)})(\delta
q_{(0,N)},\delta\lambda^{(0,N-k)})&=&
\sum_{i=0}^{N-2k}\mathcal{E}\widetilde{L}_{d}(q_{(i,2k+i)},\lambda^{(i,k+i)})\delta
q_{k+i}
\\
 &&+\Theta_{\widetilde{L}_{d}}^{+}(q_{(N-2k+1,N)})\delta q_{(N-2k+1,N)}\\
&&- \Theta_{\widetilde{L}_{d}}^{-}(q_{(0,2k-1)})\delta q_{(0,2k-1)}\\
&&+\sum_{i=0}^{N-k}\Phi_{d}^{\alpha}(q_{(i,i+k)})\delta\lambda^{i}_{\alpha}.
\end{eqnarray*}

\end{theorem}

If we consider variations at the fixed initial and final conditions
$(q_{(0,k-1)}, q_{(N-k+1,N)})$, the critical trajectories of the
unconstrained problem are given by the curves that annihilates
$\partial\widetilde{\mathcal{A}}_{d}/\partial q_{i}$ and the
constraints equations $\partial\widetilde{\mathcal{A}}_{d}/\partial
\lambda^{i}_{\alpha}.$

Thus, the \textit{higher-order discrete Euler-Lagrange equations
with constraints} are
\begin{eqnarray}\label{eqeq}
    0&=&\mathcal{E}\widetilde{L}(q_{(i,2k+i)},\lambda^{(i,N-k+i-1)})\,\qquad  0\leq i\leq N-2k \nonumber\\
    0&=&\Phi_{d}^{\alpha}(q_{(i,i+k)}) \qquad\qquad\qquad 0\leq i\leq N-k.
\end{eqnarray}

Therefore, using the implicit function theorem, we can establish the
following regularity condition (see \cite{benitodiego} for a similar proof)
\begin{proposition}\label{regularidadcondition}
If the matrix
$$\left(
       \begin{array}{cc}
         D_{(1,k+1)}L_{d}(q_{(1,k+1)})+\lambda_{\alpha}D_{(1,k+1)}\Phi_{d}^{\alpha}(q_{(1,k+1)}) & D_{k+1}\Phi_{d}^{\alpha}(q_{(1,k+1)}) \\
         (D_{1}\Phi_{d}^{\alpha}(q_{(1,k+1)})^{T} & 0 \\
       \end{array}
     \right)$$ is non-singular,
     there exists an application
      $\widetilde{\Upsilon}_{d}:\widetilde{\mathcal{M}}_{d}\times\mathbb{R}^{km}\rightarrow \widetilde{\mathcal{M}}_{d}\times\mathbb{R}^{km}$
      given by
 \begin{equation}\label{algor}\widetilde{\Upsilon}_{d}(q_{(i,i+2k-1)},\lambda^{(i,i+k-1)}):=(q_{(i+1,i+2k)},
      \lambda^{(i+1,i+k)})\end{equation}
where $q_{2k+i}$ and $\lambda_{\alpha}^{i+k}$ with
$1\leq\alpha\leq m$ is the unique solution of the equation
(\ref{eqeq})
       with initial conditions
       $(q_{(i,i+2k-1)},\lambda^{(i,i+k-1)})$
       with $0\leq i\leq N-k$.

Here, if $F$ is a smooth function on $Q^{k+1}$, $D_{(1,k+1)}F$
denotes the the partial derivative of $F$ with respect to first and
the last variables.

\end{proposition}

\begin{remark}{\textbf{Discrete Poincar\'e-Cartan 2-form:}}\label{remma}
 It is easy to shown that
\begin{equation}\label{leoleo}
\sum_{i=0}^{k-1}d\widetilde{L}_{d}(q_{(i,i+k)},\lambda^i) =
\Theta_{\widetilde{L}_{d}}^{+}(q_{(0,2k-1)},\lambda^{(0,k-1)})-\Theta_{\widetilde{L}_{d}}^{-}(q_{(0,2k-1)},\lambda^{(0,k-1)}).
\end{equation}
Therefore, using $d^{2}=0,$ it follows that
$d\Theta_{\widetilde{L}_{d}}^{-}=d\Theta_{\widetilde{L}_{d}}^{+}.$
Thus, there exists a unique 2-form $\Omega_{\widetilde{L}_{d}}:=
-d\Theta_{\widetilde{L}_{d}}^{-} =
-d\Theta_{\widetilde{L}_{d}}^{+},$ which will be called the
\textit{Discrete Poincar\'e-Cartan} 2-form.
\end{remark}

\begin{remark}{\textbf{Symplectic behavior:}} By considering the canonical inclusion
$j:\widetilde{\mathcal{M}}_{d}\times\mathbb{R}^{km}\rightarrow
Q^{k+1}\times\mathbb{R}^{km}$ we derive a 2-form
$\Omega_{\widetilde{\mathcal{M}_{d}}}:=j^{*}\Omega_{\widetilde{L}_{d}}$
on $\widetilde{\mathcal{M}}_{d}\times\mathbb{R}^{km}$ where
$\Omega_{\widetilde{L}_{d}}$ is 2-form defined on Remark
\ref{remma}. Therefore it is a natural question to ask about
conditions that ensure the symplectic character of the 2-form
$\Omega_{\widetilde{\mathcal{M}}_{d}}$. By using similar
techniques that in \cite{benitodiego} one could  establish
conditions that guarantee that the 2-form
$\Omega_{\widetilde{M}_{d}}$ is symplectic and moreover
$$(\Upsilon\mid_{\widetilde{\mathcal{M}}_{d}\times\mathbb{R}^{km}})^{*}\Omega_{\widetilde{\mathcal{M}}_{d}}=\Omega_{\widetilde{\mathcal{M}}_{d}}.$$

More specifically, if the matrix
$(D_{(1,k+1)}L_{d}+\lambda_{\alpha}D_{(1,k+1)}\Phi_{d}^{\alpha})$ is
non-singular, the discrete 2-form
$\Omega_{\widetilde{\mathcal{M}}_{d}}$ is symplectic if and only if
the matrix
$$\left(
       \begin{array}{cc}
         (D_{(1,k+1)}L_{d}+\lambda_{\alpha}D_{(1,k+1)}\Phi_{d}^{\alpha}) & D_{k+1}\Phi_{d}^{\alpha} \\
         (D_{1}\Phi_{d}^{\alpha})^{T} & 0 \\
       \end{array}
     \right)$$ is nondegenerate.
\end{remark}

\begin{remark}{\textbf{Momentum preservation:}}
Given an action of a Lie group $G$ on $Q$, we can consider the
associated $G$-action on $Q^{k+1}$ defined as
$$ g  \cdot q_{(i,k+i)} = ( g \cdot q_i, g \cdot q_{i+1}, ... , g \cdot
q_{i+k})$$ and its trivial extension on
$Q^{2k}\times\mathbb{R}^{km}$ for $g\in G.$

As this last action results symplectic, denoting by $\mathfrak{g}$
the Lie algebra associated with the Lie group $G,$ we can define two
higher-order discrete momentum maps
$$J_{d}^{\pm}:Q^{2k}\times\mathbb{R}^{km}\rightarrow\mathfrak{g}^{*}$$
given by
\begin{eqnarray*}
J_{d}^{\pm}(q_{(i,2k+i-1)},\lambda^{(i,i+k-1)})&:&\mathfrak{g}\rightarrow\mathbb{R}\\
&&\xi\mapsto\langle\Theta_{\widetilde{L}_{d}}^{\pm}(q_{(i,i+2k-1)},\lambda^{(i,i+k-1)}),\xi_{Q^{2k}}(q_{(i,i+2k-1)})\rangle,
\end{eqnarray*} for $\xi\in\mathfrak{g}.$

Is easy see that if the discrete Lagrangian $L_d$ and the discrete
constraints $\Phi_{d}^{\alpha}$ are $G$-invariant, the
higher-order discrete momentum maps coincides and then, we can
define the higher-ordern discrete momentum map that results
conserved by the discrete flow $\Upsilon_{d}$. That is,

 $$J_{d}:=J_{d}^{+}=J_{d}^{-} \ \ \ \mbox{and} \ \ \
 J_{d}\circ\Upsilon_{d}=J_{d}.$$
\end{remark}

\section{Theoretical Examples and Applications}\label{section5}

\subsection{Higher-order discrete time-dependent Lagrangian systems}
In this subsection we consider higher-order discrete
time-dependent lagrangian system with higher-order constraints.
The configuration space for this type of systems is
$\widetilde{Q}=\mathbb{R}\times Q$ where $Q$ is a $n$-dimensional
manifold. The algorithm \eqref{algor} can be adapted for obtain a
variational integrator for this kind of systems. In this case, the
discrete action $\mathcal{A}_d : \widetilde{Q}^{N+1} \rightarrow
\mathbb{R}$ is defined as
\begin{equation}\label{actiontd}
\mathcal{A}_d(t_{(0,N)},q_{(0,N)}):=\sum_{i=0}^{N-k}(t_{i+k}-t_i)L_d(t_{(i,i+k)},q_{(i,i+k)}).
\end{equation}
As it is well known (see \cite{Kane}, \cite{LD1} and references
therein) the evolution of the energy is given by the discrete
Euler-Lagrange equations corresponding to the temporal variable. The
equations involving derivatives on  $t_i$ are
\begin{eqnarray}\label{eqt}
0&=&\sum_{j=1}^{k+1}D_jL_d(t_{(i-j+1,i-j+k+1)},q_{(i-j+1,i-j+k+1)})(t_{i-j+k+1}-t_{i-j+1})\\
&&-L_d(t_{(i,i+k)},q_{(i,i+k)})+L_d(t_{(i-k,i)},q_{(i-k,i)}).\nonumber
\end{eqnarray}

By considering $h_k=t_{k+1}-t_k,$ we can define the new Lagrangian
$\overline{L}_d$ given by
\begin{eqnarray*}
&&\overline{L}_d(t_i, h_{(i,
i+k-1)},q_{(i,i+k)})=L_d(t_{(i,i+k)},q_{(i,
i+k)})=\\
&&L_{d}(t_i,t_i+h_i,t_i+h_i+h_{i+1},\ldots,t_{i}+h_i+h_{i+1}+\ldots+h_{i+k-1},q_{(i,i+k)}).
\end{eqnarray*}
Then we have the following relation between the derivatives of
$L_d$ and $\overline{L}_d$

\begin{eqnarray*}
\frac{\partial L_d}{\partial
t_j}&=&\frac{\partial\bar{L}_d}{\partial t_j} \ \ \ \hbox{ for } 1\leq j\leq k \\
D_jL_d&=&D_j\bar{L}_d-D_{j+1}\bar{L}_d \ \ \ \mbox{ for } 1\leq j\leq k\\
D_jL_d&=&D_j\overline{L}_d \ \ \ \hbox{ for } j= k+1,
\end{eqnarray*}
Substituting these expressions in \eqref{eqt} we obtain the
following equation
\begin{eqnarray*}
&&\sum_{i=1}^{k}D_{j}\overline{L}_d(t_{i-j+1},h_{(i-j+1,i-j+k)},q_{(i-j+1,i-j+1+k)})(h_{i-j+1}+\ldots+h_{i-j+k})\\
&&-D_{j+1}\overline{L}_d(t_{i-j+1},h_{(i-j+1,i-j+k)},q_{(i-j+1,i-j+k)})(h_{i-j+1}+\ldots+h_{i-j+k})\\
&&+D_{k+1}\overline{L}_d(t_{i-k},h_{(i-k,i-1)},q_{(i-k,i)})(h_{i-j+1}+\ldots+h_{i-1})-\overline{L}_{d}(t_i,
h_{(i,i+k-1)},q_{(i,i+k)})\\
&&+\overline{L}_d(t_{i-k},h_{(i-k, i-1)},q_{(i+k,i-1)})=0.
\end{eqnarray*}

The \textit{higher-order discrete energy} is defined as
\begin{eqnarray*}
E_d&=&-\frac{\partial}{\partial
h_i}\left(\sum_{j=1}^{k}\overline{L}_d(t_{i-j+1},
h_{(i-j+1,i-j+k)},q_{(i-j+1,i-j+k)})(h_{i-j+1}+\ldots+h_{i-j+k})\right)\\
&=&-\sum_{j=1}^{k}D_{j+1}\overline{L}_d(t_{i-j+1},h_{(i-j+1,i-j+k)},q_{(i-j+1,i-j+k)})(h_{i-j+1}+\ldots+h_{i-j+k})\\
&-&\sum_{j=1}^{k}\overline{L}_d(t_{i-j+1},h_{(i-j+1,i-j+k)},q_{(i-j+1,i-j+k)}).
\end{eqnarray*}
A direct computation shows that
\begin{eqnarray*}&&E_d(t_{i-k+1},h_{i-k+1},\ldots,h_{i-1},q_{(i-k+1,i-1)})-E_d(t_{i-k},h_{i-k},\ldots,h_{i+k-2},q_{(i-k,i+k-2)})=\\
&&-D_1\overline{L}_d(t_i,h_{(i,i+k)},q_{(i,i+k)})(h_i+\ldots+h_{i+k}).
\end{eqnarray*}
That is,
\begin{small}\begin{eqnarray*}
D_1\overline{L}_d(t_i,h_{(i,i+k)},q_{(i,i+k)})&=&-\frac{1}{h_i+\ldots+h_{i+k}}(E_d(t_{i-k+1},h_{i-k+1},\ldots,h_{i-1},q_{(i-k+1,i-1)})\\
&&-E_d(t_{i-k},h_{i-k},\ldots,h_{i+k-2},q_{(i-k,i+k-2)})).
\end{eqnarray*}
\end{small}
If the discrete Lagrangian is autonomous then we obtain the preservation of the discrete energy $E_d$ and the derived variational method will be a  symplectic energy- momentum
preserving method (see  \cite{Kane,LD1} for first order
systems).

\subsection{Time-dependent higher-order Lagrangians with fixed time-step size}

In the following, we consider a time-dependent Lagrangian systems
given by a lagrangian $L:\mathbb{R}\times T^{(2)}Q\ra\R$ with local
coordinates $(t,q^{A},\dot{q}^{A},\ddot{q}^{A});$ $1\leq A\leq
n=\dim Q$.  Assume for simplicity that $Q$ is a vector space This kind of systems are unconstrained, but with fixed
time step size $t_{k+1}-t_k=h$ for $k=0,\ldots,N-1,$ and $h>0.$

We may construct a discrete Lagrangian $L_d:3\R\times
3Q\rightarrow\R$ as
\begin{eqnarray*}
&&L_d(t_k,t_{k+1},t_{k+2},q_k,q_{k+1},q_{k+2})=\\
&&L\left(\frac{t_{k+2}+t_{k+1}+t_k}{3},
\frac{q_k+q_{k+1}+q_{k+2}}{3},\frac{q_{k+2}-q_k}{t_{k+2}-t_k},
\frac{\frac{q_{k+2}-q_{k+1}}{(t_{k+2}-t_{k+1})}-\frac{q_{k+1}-q_k}{(t_{k+1}-t_k)}}{(t_{k+2}-t_{k+1})}\right),\end{eqnarray*}
where $3\R\times 3Q=\R\times\R\times\R\times Q\times Q\times Q.$

Define the constraint submanifold
$$\mathcal{N}_d=\{(t_0,t_1,t_2,q_0,q_1,q_2)\in3\R\times 3Q\mid
t_1=t_0+h\hbox{ and }t_2=t_1+h\}$$ for some constant $h>0.$ This
submanifold corresponds to the vanishing of the constraints
\begin{eqnarray*}\Phi_d^{(1)}(t_0,t_1,t_2,q_0,q_1,
q_2)&=&t_1-t_0-h;\\
\Phi_d^{(2)}(t_0,t_1,t_2,q_0,q_1, q_2)&=&t_2-t_1-h, \end{eqnarray*}
and now take the augmented Lagrangian
\begin{eqnarray*}
\tilde{L}_d(t_k,t_{k+1},t_{k+2},q_k,q_{k+1},q_{k+2},\lambda)&=&L_d(t_k,t_{k+1},t_{k+2},q_k,q_{k+1},q_{k+2})\\
&+&\lambda_{\alpha}\Phi_{d}^{\alpha}(t_k,t_{k+1},t_{k+2},q_k,q_{k+1},q_{k+2}),
\end{eqnarray*} with $\alpha=1,2.$

This lagrangian gives rise to the following  equations of motion
\begin{eqnarray*}
0&=&(t_{k+2}-t_k)D_4L_d(t_k,t_{k+1},t_{k+2},q_k,q_{k+1},q_{k+2})\\
&+&(t_{k+1}-t_{k-1})D_5L_d(t_{k-1},t_k,t_{k+1},q_{k-1},q_{k},q_{k+1})\\
&+&(t_{k}-t_{k-2})D_6L_d(t_{k-2},t_{k-1},t_k,q_{k-2},q_{k-1},q_k)\\
0&=&(t_{k+2}-t_k)D_1L_d(t_k,t_{k+1},t_{k+2},q_k,q_{k+1},q_{k+2})+L_d(t_{k-2},t_{k-1},t_k,q_{k-2},q_{k-1},q_{k})\\
&+&(t_{k+1}-t_{k-1})D_2L_d(t_{k-1},t_k,t_{k+1},q_{k-1},q_k,q_{k+1})+\lambda_1^{k-1}-\lambda_2^{k-1}+\lambda_2^{k-2}-\lambda_1^{k}\\
&+&(t_k-t_{k-2})D_3L_d(t_{k-2},t_{k-1},t_k,q_{k-2},q_{k-1},q_k)-L_d(t_k,t_{k+1},t_{k+2},q_k,q_{k+1},q_{k+2})\\
0&=&t_{k+1}-t_k-h;\\
0&=&t_k-t_{k-1}-h, \hbox{ where } 2\leq k\leq N-2.
\end{eqnarray*}
Finally, observe that these equations are completely decoupled, so
we can choose  the first  equation. Therefore, we obtain
\begin{eqnarray*}
0&=&D_4L_d(t_{k},t_{k}+h,t_{k}+2h,q_k,q_{k+1},q_{k+2})\\
&+&D_5L_d(t_{k}-h,t_{k},t_{k}+h ,q_{k-1},q_{k},q_{k+1})\\
&+&D_6L_d(t_{k}-2h,t_{k}-h,t_{k},q_{k-2},q_{k-1},q_k),\\
\end{eqnarray*}
with $k=2,\ldots,N-2$ and $t_0,q_0,q_1,q_{N-q},q_N$ fixed points and
time.

Observe that this equation has precisely the same form as the
discrete Euler-Lagrange equations in the time-independent case.

 Finally, we remark that an extension of this setup can be used
for more sophisticated step size control, by taking the constraint
function to be (for example),
$$\Phi_d^{(1)}(t_k,t_{k+1},t_{k+2},q_k,q_{k+1},q_{k+2})=t_{k+1}-t_k-h(q_{k},q_{k+1},q_{k+2}) \ \ \ \mbox{and}$$
$$\Phi_d^{(2)}(t_k,t_{k+1},t_{k+2},q_k,q_{k+1},q_{k+2})=t_{k+2}-t_{k+1}-h(q_k,q_{k+1},q_{k+2}),$$
where $h:Q^3\ra\R,$ $h>0$ is some step size function. In this case,
$$D_j\Phi_d(t_k,t_{k+1},t_{k+2},q_k,q_{k+1},q_{k+2})=-D_{j-3}h(q_k,q_{k+1}.q_{k+2}),\quad j=4,5,6.$$This differs considerably from the
constant $h$.

\begin{example}
{\rm As an illustrative example of discrete time-dependent higher-order
mechanical system we consider a deformed elastic cylindrical beam
with both ends fixed. This example  is not time-dependent system,
but it can be modeled using a configuration bundle over a compact
subset of $\R,$ where the coordinates in the base configuration
represents every transversal section of the beam. We take, instead
of a compact subset, the whole real line as the base manifold. This
example has been also study in \cite{PrietoRoman-Roy2} in the
continuous setting. The second-order Lagrangian is given by

\begin{equation}\label{beam}
L(t,q,\dot{q},\ddot{q})=\frac{1}{2}\mu(t)\ddot{q}^2+\rho(t)q
\end{equation} where $\mu,\rho$ are differentiable functions that only depend on the coordinate $t$ and represent physical
parameters of the beam. If the beam is homogeneous, $\rho$ and $\mu$
are constants (with $\mu\neq 0$), and thus the Lagrangian density is
autonomous, that is, it does not depend explicitly on the coordinate
of the base manifold (see \cite{PrietoRoman-Roy2}
 and references therein).

The discrete lagrangian associated to \eqref{beam} defined on
$3(\R\times Q)$ is given by
\begin{eqnarray*}
L_d&=&\frac{1}{2}\mu\left(\frac{t_{k+2}+t_{k+1}+t_k}{3}\right)\left(\frac{q_{k+2}-q_{k+1}}{(t_{k+2}-t_{k+1})^{2}}-\frac{q_{k+1}-q_{k}}{(t_{k+1}-t_{k})(t_{k+2}-t_{k+1})}\right)^2\\
&&+\rho\left(\frac{t_{k+2}+t_{k+1}+t_k}{3}\right)\frac{q_{k+2}+q_{k+1}+q_k}{3},
\end{eqnarray*}

and the associated implicit discrete algorithm is given by
\begin{eqnarray*}
0&=&\frac{1}{3}\left[\rho\left(\triangle[t_{k+1}]\right)(h_{k+1}+h_{k})+\rho\left(\triangle[t_{k}]\right)(h_{k}+h_{k-1})+\rho\left(\triangle[t_{k-1}](h_{k-1}+h_{k-2})\right)\right]\\
&&+\mu\left(\triangle[t_{k+1}]\right)\left(\frac{q_{k+2}-q_{k+1}}{h_{k+1}^{2}}-\frac{q_{k+1}-q_{k}}{h_kh_{k+1}}\right)\frac{h_{k+1}+h_k}{h_{k+1}h_k}\\
&&-\mu\left(\triangle[t_k]\right)\left(\frac{q_{k+1}-q_{k}}{h_k^{2}}-\frac{q_{k}-q_{k-1}}{h_{k-1}h_k}\right)\left(\frac{h_k+h_{k-1}}{h_k^{2}}+\frac{h_{k}+h_{k-1}}{h_{k-1}h_k}\right)\\
&&+\mu\left(\triangle[t_{k-1}]\right)\left(\frac{q_{k}-q_{k-1}}{h_{k-1}^{2}}-\frac{q_{k-1}-q_{k-2}}{h_{k-2}h_{k-1}}\right)\frac{h_{k-1}+h_{k-2}}{h_{k-1}^{2}},\\
0&=&\frac{h_{k+1}+h_k}{3}\partial_{t_k}\rho(\partial\triangle[t_{k-1}])\triangle[q_{k-1}]-\rho(\triangle[t_{k+1}])\triangle[q_{k+1}]\\
&&+\rho(\triangle[t_{k-1}])\triangle[q_{k-1}]+\frac{1}{2}\mu(\triangle[t_{k-1}])\left(\frac{q_k-q_{k-1}}{h_{k-1}^2}-\frac{q_{k-1}-q_{k-2}}{h_{k-2}h_{k-1}}\right)^{2}\\
&&+\frac{h_{k+1}+h_{k}}{6}\partial_{t_k}\mu(\triangle[t_{k-1}])\left(\frac{q_{k+2}-q_{k+1}}{h_{k+1}^2}-\frac{q_{k+1}-q_k}{h_k h_{k+1}}\right)^{2}\\
&&-\frac{1}{2}\mu(\triangle[t_{k-1}])\left(\frac{q_k-q_{k+1}}{h_{k-1}^2}-\frac{q_{k+1}-q_k}{h_{k-2}h_{k-1}}\right)^2+\frac{h_k+h_{k-1}}{3}\partial_{t_k}\mu(\triangle[t_k])\triangle[q_k]\\
&&-\frac{(q_{k+1}-q_k)(h_{k+1}+h_k)}{h_{k}^2h_{k+1}^2}\mu(\triangle[t_{k-1}])\left(\frac{q_{k+2}-q_{k+1}}{h_{k+1}^2}-\frac{q_{k+1}-q_{k}}{h_{k}h_{k+1}}\right)\\
&&+\frac{h_k+h_{k-1}}{6}\partial_{t_k}\mu_{t_k}\mu(\triangle[t_k])\left(\frac{q_{k+1}-q_{k}}{h_{k}^{2}}-\frac{q_k-q_{k-1}}{h_{k-1}h_{k}}\right)^{2}\\
&&+\frac{h_{k-1}+h_{k-2}}{3}\partial_{t_k}\rho(\triangle[t_{k-1}])\triangle[q_{k-1}]\\
&&+h_{k}\mu(\triangle[t_k])\left(\frac{q_{k+1}-q_k}{h_{k}^{2}}-\frac{q_{k}-q_{k-1}}{h_{k-1}h_{k}}\right)\left(\frac{2(q_{k+1}-q_{k})}{h_{k}^{3}}-\frac{(q_{k}-q_{k-1})(h_{k}-h_{k+1})}{h_{k-1}^{2}h_{k}^{2}}\right)\\
&&+\frac{h_{k-1}+h_{k-2}}{6}\partial_{t_k}\mu(\triangle[t_{k-1}])\left(\frac{q_{k}-q_{k-1}}{h_{k-1}^{2}}-\frac{q_{k-1}-q_{k-2}}{h_{k-2}h_{k-1}}\right)^{2}\\
&&-\mu(\triangle[t_{k-1}])\left(\frac{q_k-q_{k-1}}{h_{k-1}^2}-\frac{q_{k-1}-q_{k-2}}{h_{k-2}h_{k-1}}\right)\left(\frac{2(q_{k}-q_{k-1})}{h_{k-1}^{3}}-\frac{q_{k-1}-q_{k-2}}{h_{k-2}h_{k-1}^{2}}\right),
\end{eqnarray*}  for $2\leq k\leq
N-2$ where $h_{k}=t_{k+1}-t_{k},$
$\triangle[t_k]=\frac{t_{k+1}+t_k+t_{k-1}}{3},$
$\triangle[q_k]=\frac{q_{k+1}+q_k+q_{k-1}}{3}$ and $\partial_{t_k}$
denotes the partial derivative of a function with respect to the
variable $t_k.$

}
\end{example}

\subsection{Optimal control of underactuated time-dependent
menchanical systems}

In this subsection, we will construct a variational integrator for the
time-dependent underactuated optimal control problem that we have introduced in Subsection \ref{asw}.

Consider a discrete second-order time-dependent Lagrangian
system  given by the function $L_d:(\R\times Q)^2\rightarrow \R$
where $Q=Q_1\times Q_2.$ An element $(t_0,q_0^i,t_1, q_1^i)\in
(\R\times Q)^2$ admits a global decomposition of the form
$(t_0,q_0^a, q_0^\alpha,t_1, q_1^a, q_1^\alpha)$ with $1\leq a\leq
m$, $m+1\leq \alpha\leq n$  and the discrete second-order
constraints are given by $\Phi_{d}^{\alpha}:(\R\times
Q)^2\rightarrow \R,$ determining the submanifold
$\mathcal{M}_{d}$.


Consider the following \textit{discrete time-dependent underactuated
mechanical  system,}
\begin{eqnarray*}\label{aqr}
(t_i-t_{i-1})D_4^aL_{d}(t_{i-1},q_{i-1}^{A},t_i, q_{i}^{A}) +
(t_{i+1}-t_i)D_2^aL_{d}(t_i,q_{i}^{A},
t_i,q_{i+1}^{A}) &=& u_i^{a}\\
(t_i-t_{i-1})D_4^{\alpha}L_{d}(t_{i-1},q_{i-1}^{A},t_i, q_{i}^{A}) + (t_{i+1}-t_i)D_2^{\alpha}L_{d}(t_i,q_{i}^{A},t_i, q_{i+1}^{A}) &=& 0\\
\end{eqnarray*}
with $1\leq i\leq n$, $1\leq a\leq m$ and $m+1\leq \alpha\leq n$.
Denote by $D^{a}_i$ and $ D^{\alpha}_i$  the partial
derivatives with respect to coordinates $a$ and $\alpha$,
respectively.

The optimal control problem is determined prescribing the discrete
cost functional
$$\mathcal{A}_{d}(t_{(0,N)},q^A_{(0,N)},u^{\alpha}_{(0,N-1)}) =
\sum_{i=0}^{N-1} C(t_i,q^A_{i},t_{i+1}, q^A_{i+1}, u_i^{a})$$ with
initial and final conditions $t_0,q_0,t_1,q_1$ and
$t_{N-1},q_{N-1},t_N,q_N$ respectively.

Since the control variables appear explicitly the previous optimal
control problem is equivalent to the second-order variational
problem with constraints determined by
$$\min\widetilde{\mathcal{A}}_d(t_{(0,N)},q^A_{(0,N)},u^{\alpha}_{(0,N-1)})=\sum_{i=0}^{N-2}\widetilde{L}_d(t_{i}, q_{i}^{A},t_{i+1}, q_{i+1}^{A},t_{i+2}, q_{i+2}^{A})$$ and the  constraints
\begin{eqnarray*}
&&\Phi_d^\alpha(t_{i}, q_{i}^{A},t_{i+1}, q_{i+1}^{A},t_{i+2},
q_{i+2}^{A})=\\
&&(t_{i+1}-t_{i})D_4^{\alpha}L_{d}(t_{i},q_{i}^{A},t_{i+1},
q_{i+1}^{A}) +
(t_{i+2}-t_{i+1})D_2^{\alpha}L_{d}(t_{i+1},q_{i+1}^{A},t_{i+2},
q_{i+2}^{A})=0
\end{eqnarray*}
where,
\begin{eqnarray*}
\widetilde{L}_{d}(t_{i},q_{i}^{A},t_{i+1}, q_{i+1}^{A},t_{i+2}, q_{i+2}^{A}) &=& C\left(t_{i},q_{i}^{A},t_{i+1},
q_{i+1}^A,(t_{i+1}-t_{i})D_4^aL_{d}(t_{i},q_{i}^{A},t_{i+1}, q_{i+1}^{A}), \right.\\
&&+\left.
(t_{i+2}-t_{i+1})D_2^aL_{d}(t_{i+1},q_{i+1}^{A},t_{i+2},
q_{i+2}^{A})\right).
\end{eqnarray*}

Now, define $\overline L_{d}:(\R\times
Q)^3\times\R^{m}\rightarrow\R$ by $\overline
L_{d}=\widetilde{L}+\lambda_{\alpha}\Phi_d^{\alpha}$
and our problem is related to the discrete variational problem  $$\min
\overline
{\mathcal{A}}_{d}(t_{(0,N)},q^a_{(0,N)},q^{\alpha}_{(0,N)},
\lambda_{\alpha}^{(0,N-2)})$$

 where
 \begin{eqnarray*}
&&\overline{\mathcal{A}}_{d}(t_{(0,N)},q^a_{(0,N)},q^{\alpha}_{(0,N)},
\lambda_{\alpha}^{(0,N-2)}) =\\
&&\sum_{i=0}^{N-2}\overline
L_{d}(t_{i},q_{i}^a,q_i^{\alpha},t_{i+1},
q_{i+1}^a,q_{i+1}^{\alpha}, t_{i+2}, q_{i+2}^a, q_{i+2}^{\alpha},
\lambda_{\alpha}^i).
\end{eqnarray*}

In order to apply the techniques developed in the previous section
(where the configuration space is $\R\times Q)$ we assume the
regularity condition given in Theorem \ref{regularidadcondition}.

Thus, for all point in $\mathcal{M}_d=\{(r,x,s,y,t,z)\in(\R\times
Q)^3\mid \Phi_{d}^{\alpha}(r,x,s,y,t,z)=0\}$ and
$\lambda_{\alpha}\in\R^{m} \hbox{ with } 1\leq\alpha\leq m,$  the
discrete flow
\[
\begin{array}{rrcl}
\Upsilon_d:&\overline{\mathcal M}_d\times \R^{2m}&\longrightarrow&\overline {\mathcal M}_d\times \R^{2m}\\
    & (t_0, q_0, t_1, q_1, t_2, q_2, t_3, q_3,\lambda^0_{\alpha}, \lambda^1_{\alpha})&\longmapsto& (t_1, q_1, t_2, q_2, t_3, q_3, t_4, q_4,\lambda^1_{\alpha}, \lambda^2_{\alpha})
\end{array}
\] is given by $t_4,q_4$ and $\lambda_2,$  determined  from the initial conditions  $(t_0, q_0, t_1, q_1, t_2, q_2, t_3, q_3,\lambda^0_{\alpha}, \lambda^1_{\alpha})$.

Here, $\overline{\mathcal M}_d$ denotes the submanifold of
$(\R\times Q)^4$ given by
\begin{small}\[ \overline{\mathcal M}_d=\{ (t_0, q_0, t_1, q_1, t_2, q_2,
t_3, q_3)\; |\; \Phi^{\alpha}_d(t_0, q_0, t_1, q_1, t_2, q_2)=0,
\Phi^{\alpha}_d(t_1, q_1, t_2, q_2, t_3, q_3)=0 \}\;
\]\end{small} with $1\leq \alpha\leq m.$

Using similar techniques than in Section \ref{section4} it is
possible to show that, under the regularity assumptions, this
discrete flow is symplectic.

\subsection{Interpolation problem on Riemannian manifolds}

In what follows, we will obtain a geometric integrator for the
interpolation problem considered in Section \ref{interpolation},
but, in this case, we will add a holonomic constraint given by the
restriction to the sphere $\mathbb{S}^2$ on $\R^3.$
%
More concretely, the configuration manifold
is $Q=\R^3$ with the Euclidean metric and subject to the holonomic
constraint
$$\Phi(q)=q\cdot q-r^2=0,\qquad q\in\R^3$$ where $r>0$ is the radio of the sphere on $\R^3$ centering in the origin and $\cdot$ denotes the Euclidean inner product on $Q.$
This constraint determines the submanifold of $Q$ given by
$$\mathcal{M}=\{q \in\R^3\mid q\cdot q=r^2\}.$$

The discrete Lagrangian $L_d:3 \R^3\ra\R$   is given by
\begin{equation}\label{ld1}
L_d(q_0,q_{1},q_{2})=\frac{h}{2}\left(\frac{q_2-2q_1+q_0}{h^2}\right)^{2}
\end{equation} with $h>0$ the time step.

Fix a subset $I$ where $I \subset \{2,...,N-2\}$ representing the indices corresponding to the interpolating constraints.

Therefore, the discrete interpolating problem consists on  finding
a path $q_{(0,N)}$ minimizing the cost functional
$$\mathcal{A}_{d}(q_{(0,N)})=\sum_{k=0}^{N-2}L_d(q_k,q_{k+1},q_{k+2})$$ subject the constraint $$\Phi_d(q_k)=q_k\cdot q_k-r^2=0,$$ and fixed
the interpolating points $q_i\in \mathbb{S}^2$ for all  $i\in I$ and
$q_0,q_1,q_{N-1},q_N\in\mathbb{S}^2$ given initial and final conditions. That is,
to find a path $(q_0,q_1,\ldots,q_N)$ which solves the equations
\begin{eqnarray*}0&=&D_1L_d(q_k,q_{k+1},q_{k+2})+\lambda^{k}D\Phi_d(q_k)+D_2L_d(q_{k-1},q_{k},q_{k+1})+D_3L_d(q_{k-2},q_{k-1},q_k)\\
0&=&\Phi_d(q_{k+2})
\end{eqnarray*} for $k\in\{2,\ldots,N-2\}\backslash I$ (that is, except on points of $I$) and the interpolating constraints and the initial and final conditions.

In other words, the solution of the interpolation problem is the path which
solves the equations
\begin{eqnarray*}
0&=&\frac{1}{h^3}\left(q_{k+2}-4q_{k+1}+6q_k-4q_{k-1}+q_{k-2}\right)+2\lambda^{k}q_k \hbox{ for } k\notin I\\
0&=&q_{k+2}^2-r^2\\
q_i&=&q(t_i), \hbox{ for } i\in I
\end{eqnarray*} with $k=2,\ldots, N-2$ for paths $(q_0,q_1,q_2,\ldots,q_N)$ such
that $q_j\in \mathbb{S}^2$ with $j=0,\ldots,N$ and
$q_0,q_1,q_{N-1},q_N$ are given boundary conditions.

From these equations, we obtain the following systems of equations
\begin{eqnarray*}
\lambda^{k}&=&-\frac{1}{2r^2h^3}(q_{k+2}q_k-4q_{k+1}q_k+q_{k-2}q_k-4q_{k-1}q_k+6r^{2})\\
0&=&\frac{1}{h^3}(q_{k+2}-4q_{k+1}+6q_k-4q_{k-1}+q_{k-2})\\
&&-\frac{q_k}{r^2h^3}(q_{k+2}q_k-4q_{k+1}q_k+q_{k-2}q_k-4q_{k-1}q_k+6r^2)
\end{eqnarray*}

\subsection{Conclusions}
In this paper we have developed a variational integrator for higher-order
Lagrangian systems with constraints.
We have considered the case of time-dependent Lagrangian systems, and
we have analyzed the behavior of the energy evolution  associated with
this type of systems. Moreover, we have also studied time-dependent second order
constrained mechanics with fixed time-stepping.

We have derived variational integrators for higher-order Lagrangian mechanics
with constraints in some interesting cases, for instance, an optimal control problem for an
underactuated time-dependent mechanical systems and an interpolation
problem for Riemannian manifolds.

\end{document}